\documentclass[regular]{jfm}

\usepackage{graphicx}
\usepackage{newtxtext}
\usepackage{newtxmath}
\usepackage{natbib}
\usepackage{hyperref}
\usepackage{bm}
\usepackage{longtable}
\usepackage{booktabs}            %插表格用的宏包
\usepackage{diagbox}             %插表格用的宏包
\usepackage{multirow}            %插多行表格用的宏包
\hypersetup{
	colorlinks = true,
	urlcolor   = blue,
	citecolor  = blue,
	linkcolor  = blue
}

\newcommand\Ren{\mbox{\textit{Re}}}
\newcommand\Ra{\mbox{\textit{Ra}}}

\newcommand\Nun{\mbox{\textit{Nu}}}
\newcommand\Prn{\mbox{\textit{Pr}}}

\newcommand\Am{\mbox{\textit{a}}}
\usepackage[usenames,dvipsnames]{xcolor}
\shorttitle{Unifying constitutive law of vibroconvective turbulence in microgravity}
\shortauthor{J.-Z. Wu et al.}
\title{Unifying constitutive law of vibroconvective turbulence in microgravity}

\author{Ze-Lin Huang, 
    Jian-Zhao Wu \corresp{\email{jianzhao\_wu@shu.edu.cn}},
	Xi-Li Guo, 
	Chao-Ben Zhao,
	Bo-Fu Wang, 
	Kai Leong Chong,
	\and Quan Zhou\corresp{\email{qzhou@shu.edu.cn}}
}

\affiliation{Shanghai Key Laboratory of Mechanics in Energy Engineering, Shanghai Institute of Applied Mathematics and Mechanics, School of Mechanics and Engineering Science, Shanghai University, Shanghai 200072, China}

\begin{document}
	\maketitle
	
	\begin{abstract}
        We report the unified constitutive law of vibroconvective turbulence in microgravity, i.e., $Nu \sim a^{-1} Re_\mathrm{os}^\beta$ where the Nusselt number $Nu$ measures the global heat transport, $a$ is the dimensionless vibration amplitude, $Re_\mathrm{os}$ is the oscillational Reynolds number, and $\beta$ is the universal exponent.
We find that the dynamics of boundary layers plays an essential role in vibroconvective heat transport and the $Nu$-scaling exponent $\beta$ is determined by the competition between the thermal boundary layer (TBL) and vibration-induced oscillating boundary layer (OBL). 
Then a physical model is proposed to explain the change of scaling exponent from $\beta=2$ in the TBL-dominant regime to $\beta = 4/3$ in the OBL-dominant regime. We conclude that vibroconvective turbulence in microgravity defines a distinct universality class of convective turbulence under the normal gravity field.
Our finding elucidates the emergence of universal constitutive laws in vibroconvective turbulence, and opens up a new avenue for generating a controllable effective heat transport under microgravity or even microfluidic environment in which gravity effect is nearly absent.
	\end{abstract}
		
	
	{\bf MSC Codes }  {\it(Optional)} Please enter your MSC Codes here
	
\section{Introduction}
	\label{intro}
The emergence of unified constitutive law is a hallmark of gravity-induced convective turbulence \citep{Niemela2000Nature,Grossmann2000JFM,Grossmann2001PRL,Ahlers2009RMP,Lohse2010ARFM,Stevens2013JFM,Zhu2018NP,Sreenivasan2019PNAS,Wang2019NC,Jiang2020SA}, e.g., $Nu \sim Ra^\beta$ with $\beta \approx 0.3$ in the classical regime \citep{Malkus1954PRSLA,Priestley1954AJP,Castaing1989JFM,Plumley2019Earth,Iyer2020PNAS,Ahlers2022aspect} and $\beta=1/2$ in the ultimate regime for paradigmatic Rayleigh-B{\'e}nard (RB) convection \citep{Kraichnan1962PoF,Spiegel1963AstJ,He2012PRL,Toppaladoddi2017PRL,Lepot2018PNAS,zou2021jfm}, where the Nusselt number $Nu$ quantifies the heat transport efficiency and the Rayleigh number $Ra$ quantifies the strength of buoyancy forcing. However, in microgravity, as the gravity effect is however almost absent, gravity-induced convection becomes too feeble to transport matter and heat. Vibration, omnipresent in science and technology, has been shown to be an attractive way to operate fluids, modulate convective patterns, and control heat transport by creating an ``artificial gravity'' \citep{Beysens2006EuroNews,Beysens2005PRL}, e.g., vibration shapes liquid interfaces in arbitrary direction \citep{Sanchez2020JFM,Sanchez2019JFM,Apffel2021PNAS}, vibration levitates a fluid layer upon a gas layer \citep{Apffel2020nature}, vibration selects patterns through the parametric response \citep{Rogers2000PRL,Rogers2000PRLb,Pesch2008JFM,Salgado2019PRE}, vibration significantly enhances or suppresses heat transport depending on the mutual direction of vibration and temperature gradient \citep{Wang2020SA,Swaminathan2018JASA,Wu2021PoF,wu2022jfm}. Vibroconvection, resulting directly from a non-isothermal fluid subjected to the external vibration, is very pronounced under microgravity conditions and provides a potential mechanism of heat and mass transport in absence of gravity-induced convection \citep{Gershuni1998WS,Mialdun2008PRL,Shevtsova2010JFM}. Elucidating the potential constitutive law of vibroconvective turbulence and its underlying mechanism is not only of great importance in microgravity science, but also provides practical guiding significance for space missions \citep{Monti2001AA} and microfluidic technologies \citep{Daniel2005Langmuir,Brunet2007PRL}. 

In past decades, due to the difficulty of microgravity experiments, the experimental studies on vibroconvection at low gravity were limited. A known experiment was carried out with the ALICE-2 instrument onboard MIR station, which revealed the vibrational influence on the propagation of a temperature wave from a heat source in near-critical fluids \citep{Zyuzgin2001CS,Garrabos2007PRE}. The other known experiment was conducted in the parabolic flights during the 46-th campaign organized by the European Space Agency, which reported the first direct experimental evidence of vibroconvection in low gravity \citep{Mialdun2008PRL,Shevtsova2010JFM}. 
There are extensive theoretical and numerical investigations of vibroconvection under weightlessness conditions. In the limiting case of high-frequencies and small amplitudes, the averaging technique was applied to theoretically deduce the dynamical equation of the mean flows \citep{Gershuni1998WS}. Based on the averaged equations, the onset and bifurcation scenarios of vibroconvection were widely investigated in square, rectangular, and cubic enclosures \citep{Savino1998CF,Cisse2004IJHMT}. The synchronous, subharmonic and non-periodic responses to external vibration were observed in vibroconvection from a parametric study over a wide range of frequencies and amplitudes \citep{Hirata2001JFM,Crewdson2021IJT}. The parametric and Rayleigh-vibrational instability were examined in vibroconvection in the absence of gravity \citep{Amiroudine2008PRE,Sharma2019PRF}. The heat transport enhancement near the onset of vibroconvection were also investigated \citep{Gershuni1998WS,Shevtsova2010JFM}. However, the basic properties of constitutive law in vibroconvective turbulence have been rarely addressed.

In this paper, we carried out a series of direct numerical simulations on vibroconvection in a wide range of vibration amplitudes and frequencies. Then we theoretically and numerically unveil the emergence of unified constitutive law and underlying mechanism of vibroconvective turbulence. In \S \ref{sec:dns}, governing equations and numerical approach of microgravity vibroconvection are described; In \S \ref{sec:Result and discussion}, flow structure in vibroconvection is analyzed and the unified scaling law of vibroconvective heat transport is revealed therorectially and examined numerically; Finally, the conclusion is given in \S \ref{sec:conclusion}.

\section{Direct numerical simulation}
\label{sec:dns}
We consider the microgravity vibroconvection setup of the convective flows in an enclosure heated from below by a hot wall and cooled from above by a cold wall, and subjected to the harmonic vibration $A \cos(\Omega t)$ in horizontal direction (see figure~S1 in supplementary materials). Here, $\Omega$ and $A$ are the angular frequency and pulsating displacement. In the non-inertial frame associated to the imposed vibration, an inertial acceleration of $A\Omega^2 \cos(\Omega t) \boldsymbol{e}_x$ is added to the system, where $\boldsymbol{e}_x$ is the unit vector in $x$-direction. The governing equations for vibroconvective turbulence is then can be writen as
\begin{eqnarray}
&\partial_t \boldsymbol{u} + (\boldsymbol{u} \cdot \nabla) \boldsymbol{u} = -{\nabla}{p} + \nu \nabla^2 \boldsymbol{u} - \alpha  A \Omega^2 \cos(\Omega t) T \boldsymbol{e}_{x}, \\
&\partial_t T + (\boldsymbol{u} \cdot \nabla) T =\kappa \nabla^2 T,
\end{eqnarray}
in addition to $ \nabla \cdot \boldsymbol{u}=0$, where $\boldsymbol{u}$ is the fluid velocity, $T$ the temperature, $p$ the kinematic pressure, $\nu$ the kinematic viscosity, $\kappa$ thermal diffusivity, and $\alpha$ thermal expansion coefficient, respectively. All quantities studied below have been made dimensionless with respect to the cell's height $H$, the temperature difference across the fluid layer $\Delta$, and the viscous diffusion velocity $\nu/H$. Based on these choices, the relevant control parameters for the vibroconvection system are the dimensionless vibration amplitdue $\Am = \alpha \Delta A/H$, the dimensionless vibration frequency $\omega = \Omega H^2/\nu$, and the Prandtl number $\Prn = \nu/\kappa$.

We performed direct numerical simulation of microgravity vibroconvection in a rectangular enclosure of aspect-ratio of $W$:$D$:$H$ = $1$:$0.3$:$1$ in three-dimensional cases and of $W$:$H$=$1$:$1$ in two-dimensional cases, where $W$, $D$, $H$ are respectively the width, depth and height of convection cell.
The governing equations are numerically solved by a second-order finite difference code, which has been validated many times in the literature \citep{Wang2020SA,wu2022jfm,guo2023flow}.
At all solid boundaries, no-slip boundary conditions are applied for the velocity. At top and bottom plates, constant temperatures $\theta_\mathrm{top}=0$ and $\theta_\mathrm{bot}=1$ are given; and at all side walls, the adiabatic conditions are adopted.
We performed a series of direct numerical simulations of microgravity vibroconvective turbulence over the vibration amplitude range $0.001 \le \Am \le 0.1$ and the frequency range $10^5 \le \omega \le 10^7$ for 3D cases, and over the vibration amplitude range $0.001 \le \Am \le 0.3$ and the frequency range $10^3 \le \omega \le 10^7$ for 2D cases at fixed Prandtl number $\Prn = 4.38$.
For all simulations, the computational mesh size is chosen to adequately resolve the dynamics both the thermal and oscillating boundary layers, and the time step is chosen to not only fulfil the Courant-Friedrichs-Lewy (CFL) conditions, but also resolve the time scale of one percent of the vibration period. Further details of numerical parameters are given in supplementary materials.

\section{Results and Discussion}
\label{sec:Result and discussion}
%\subsection{Flow structure}
%\label{subsec:flowstructure}

\begin{figure}
	\centerline{\includegraphics[width=1.0\linewidth]{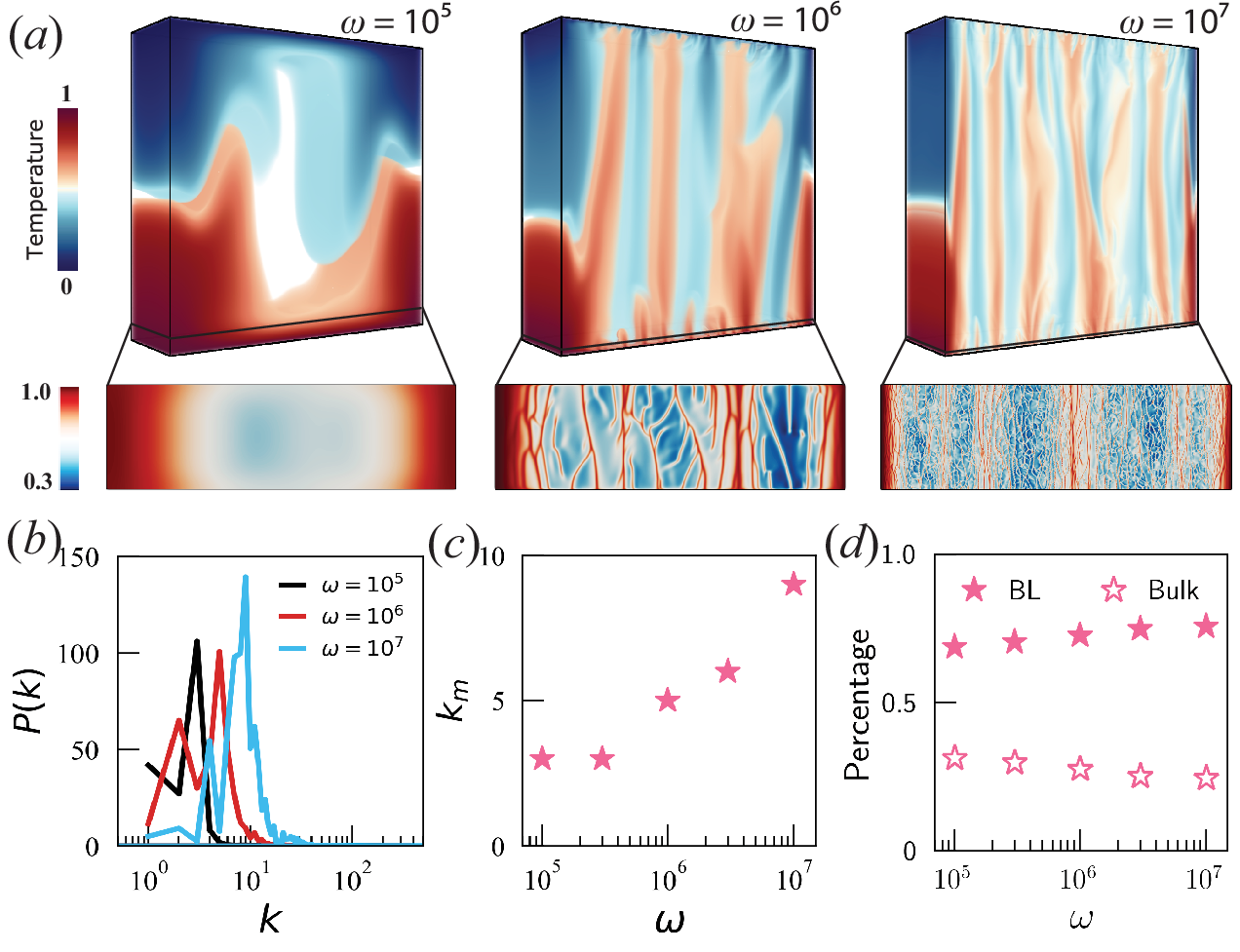}}
	%\begin{center}
	%	\includegraphics[width=0.5\linewidth]{fig1}
	%\end{center}
	\caption{{Flow structure in microgravity vibroconvection.}
		(\textit{a}) Instantaneous 3D flow structure visualized by the volume rendering of instantaneous temperature field (see supplementary movies) under different vibration frequencies $\omega = 10^5$ (left), $10^6$ (middle), and $10^7$ (right) at fixed amplitude $\Am = 0.01$ and Prandtl number $\Prn = 4.38$. Below shows the corresponding temperature contours extracted on the horizontal slice at the edge of thermal boundary layer (BL).   (\textit{b}) Power spectrum of fluctuating temperature in bulk zones. (\textit{c}) The variation of the characteristic wave number $k_m$ as functions of $\omega$. (\textit{d}) Percentage of BL (solid symbols) and bulk (hollow symbols) to the global averaged thermal dissipation rate, as functions of vibration frequency. } 
	\label{fig:flowstructure}
\end{figure}

Figure~\ref{fig:flowstructure}($a$) shows the typical snapshots of flow structures in vibroconvection with different dimensionless frequencies $\omega=10^5$, $10^6$, and $10^7$ at fixed dimensionless amplitude $\Am=0.01$ and fixed Prandtl number $\Prn = 4.38$. It is seen that the shaking by external vibration strongly destabilizes the conductive state and generates large distortion of temperature field in bulk regions by creating an artificial gravity \citep{Beysens2006EuroNews,Beysens2005PRL}. With increasing $\omega$, it is vibration-induced artificial gravity that becomes strong enough to destabilize thermal boundary layer and trigger abundant thermal plumes. Those plumes are transported into bulk regions and self-organized into columnar structures. This indicates that the feature of main structures responsible for heat transport in microgravity vibroconvection is different from that in the gravity-induced RB convection. 

To quantitatively analyze the feature of columnar structures, we extract the instantaneous temperature field in bulk zones and calculate the power spectrum $P(k)$ of temperature fluctuations by applying the Fourier transform in the vibrational direction as shown in figure~\ref{fig:flowstructure}($b$). It is found that there exists a characteristic wave number $k_{m}$, at which the wave number distribution function $P(k)$ is maximal. Indeed, $k_{m}$ characterizes the number of columnar structures in vibroconvection. We then plot the variation of $k_{m}$ as functions of $\omega$ in figure~\ref{fig:flowstructure}($c$). It is shown that $k_m$ monotonically increases with increasing $\omega$, indicating that more columnar structures are formed under stronger vibrational driving force. This is consistent with the fact that larger heat transport enhancement occurs at larger $\omega$. Further, to examine the role of thermal boundary layer in vibroconvective heat transport processes, we decompose the globally averaged thermal dissipation rate $\epsilon_T = \kappa \lvert \nabla T \rvert^2$ into their boundary layer (BL) and bulk contributions,  and then plot the variation of relative contributions as functions of $\omega$ in figure~\ref{fig:flowstructure}($d$), as suggested by the Grossmann-Lohse theory \citep{Grossmann2000JFM,Grossmann2001PRL,Stevens2013JFM}.  It is seen in figure~\ref{fig:flowstructure}($d$) that the BL contribution of $\epsilon_T$ is much larger than the bulk one, suggesting the BL-dominant thermal dissipation. This reveals that the dynamics of boundary layers plays a crucial role on the underlying mechanism of heat transport in vibroconvective turbulence.

%\subsection{Heat transport scaling}
\begin{figure}\centerline{\includegraphics[width=1.0\linewidth]{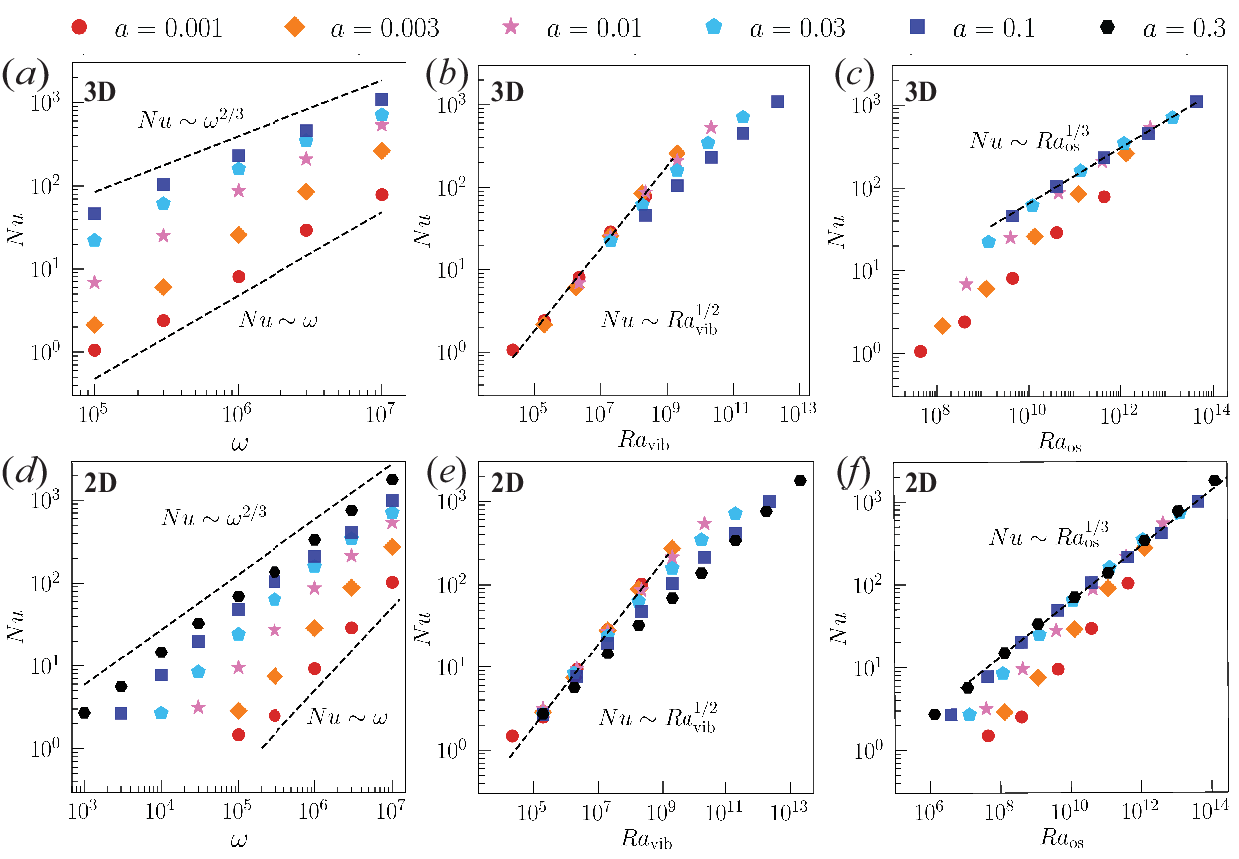}}
	%\begin{center}
	%	\includegraphics[width=0.5\linewidth]{fig1}
	%\end{center}
	\caption{{Heat transport scaling in vibroconvective turbulence.}
		(\textit{a}-\textit{c}) The measured Nusselt number $\Nun$ as functions of vibration frequency $\omega$, vibrational Rayleigh number $\Ra_\mathrm{vib}$, oscillational Rayleigh number $\Ra_\mathrm{os}$ for three-dimensional (3D) cases. 
		(\textit{d}-\textit{f}) The measured Nusselt number $\Nun$ as functions of vibration frequency $\omega$, vibrational Rayleigh number $\Ra_\mathrm{vib}$, oscillational Rayleigh number $\Ra_\mathrm{os}$ for two-dimensional (2D) cases. The dashed lines in figures are $\Nun \sim \omega$ (lower), $\Nun \sim \omega^{2/3}$ (upper) in (\textit{a} and \textit{d}), $\Nun \sim \Ra_\mathrm{vib}^{1/2}$ in (\textit{b} and \textit{e}), $\Nun \sim \Ra_\mathrm{os}^{1/3}$ in (\textit{c} and \textit{f}). Those precise scaling relations are theoretically deduced by our proposed physical model in the paper. Note that the vibration amplitude range in 3D cases is from $\Am=10^{-3}$ to $\Am=10^{-1}$ and in 2D cases is from $\Am=10^{-3}$ to $\Am=3 \times 10^{-1}$.} 
	\label{fig:nuscaling}
\end{figure}

Next, we address the question of how the global heat transport depends on the control parameters of vibroconvection. First, we examine the dependence of heat transport on the vibration frequency. Figures~\ref{fig:nuscaling}($a,d$) show the measured $\Nun$ as functions of frequency $\omega$ in a log-log plot for different amplitudes $\Am$ in three-dimensional (3D) and two-dimensional (2D) cases. Here, the $\Nun$ number, as the nondimensional ratio of the measured heat flux to the conductive one, is calculated by $\Nun = \langle w T - \kappa \partial_z T\rangle/(\kappa \Delta/H)$, where $w$ is the vertical velocity and $\langle \cdot \rangle$ denotes the time and space averaging. It is observed that the $\Nun-\omega$ scaling relation is not unique for a specific amplitude, namely, there seems to be a transition from $\Nun \sim \omega^{1}$ to $\Nun \sim \omega^{2/3}$ in both 3D and 2D cases, as shown by the dashed lines or in figure~S4 in supplementary materials. Note that the precise values of scaling exponents are obtained from the physical model we proposed below, not adjusted from the fitting with the numerical data.

Further, we examine the dependency of heat transport on the two important analogous Rayleigh numbers in vibroconvective turbulence, which are the vibrational Rayleigh number $\Ra_\mathrm{vib}$ and oscillational Rayleigh number $\Ra_\mathrm{os}$. The first one is the vibrational Rayleigh number $\Ra_\mathrm{vib} = (\alpha A \Omega \Delta H)^2/(2\nu\kappa)$, which is obtained from applying the averaged approach on vibroconvective equations in the limit of small amplitudes and high frequencies, and quantifies the intensity of the external vibrational source. Figures~\ref{fig:nuscaling}($b,e$) depict respectively the measured $\Nun$ as functions of $\Ra_\mathrm{vib}$ in a log-log plot at different amplitudes for 3D and 2D cases. We find that at small $\Ra_\mathrm{vib}$, numerical data almost collapse together on the same scaling law, i.e.,  $\Nun \sim \Ra_\mathrm{vib}^{1/2}$, as shown by the dashed lines. However, at large $\Ra_\mathrm{vib}$, a significant departure from this scaling behavior is observed for large amplitudes.
The other is the oscillational Rayleigh number $\Ra_\mathrm{os} = \alpha A \Omega^2 \Delta H^3/(\nu\kappa)$, which is analogous to Rayleigh number in RB convection but replacing the gravitation by the vibration-induced acceleration. Figures~\ref{fig:nuscaling}($c,f$) show the variation of $\Nun$ as functions of $\Ra_\mathrm{os}$ for various amplitudes in 3D and 2D cases. We find that at large $\Ra_\mathrm{os}$, numerical data almost collapse onto the same scaling relation $\Nun \sim \Ra_\mathrm{os}^{1/3}$ as shown by the dashed line, but at small $\Ra_\mathrm{os}$, numerical data points deviate a lot from this scaling for small amplitudes. Both $\Nun \sim \Ra_\mathrm{vib}^{1/2}$ and $\Nun \sim \Ra_\mathrm{os}^{1/3}$ show the independence of the cell height $H$, but exhibit different scaling behaviors with the intensity of vibration and the temperature difference $\Delta$ between hot and cold plates. From above, using solely the common control parameters like $\omega$, $\Ra_\mathrm{vib}$ or $\Ra_\mathrm{os}$, unifying the heat transport scaling in vibroconvective turbulence can not be achieved.
 
\begin{figure}\centerline{\includegraphics[width=1.0\linewidth]{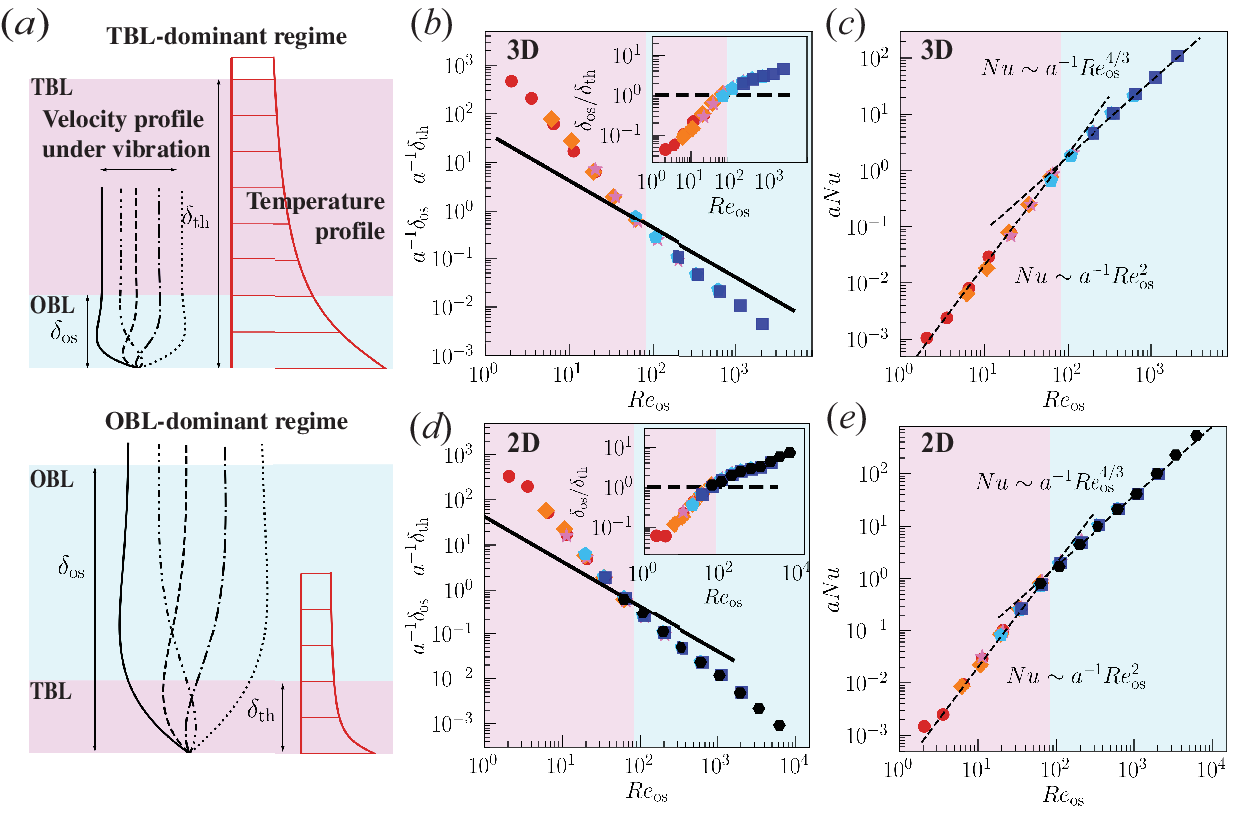}}
	%\begin{center}
	%	\includegraphics[width=0.5\linewidth]{fig1}
	%\end{center}
	\caption{{Unified constitutive law in vibroconvective turbulence.}
		(\textit{a}), Sketch of thermal-boundary-layer (TBL) dominant regime (upper) and  oscillating-boundary-layer (OBL) dominant regime (lower). 
		(\textit{b},\textit{d}) The TBL thickness $\delta_\mathrm{th}$ (symbols) and the critical modulation depth $\delta_\mathrm{os}$(solid line) of OBL as a function of the oscillational Reynolds number $\Ren_\mathrm{os}$ for 3D and 2D cases. The insets in (\textit{b},\textit{d}) show the ratio $\delta_\mathrm{os}/\delta_\mathrm{th}$ as a function of the oscillational Reynolds number $\Ren_\mathrm{os}$. The solid dashed line indicates $\delta_\mathrm{os}/\delta_\mathrm{th} = 1$. 
		(\textit{d},\textit{e}) The unified scaling law exhibited between $a \Nun$ and the the oscillational Reynolds number $\Ren_\mathrm{os}$ for 3D and 2D cases. The emergency of universal constitutive law of vibroconvective turbulence is clearly observed.} 
	\label{fig:Unifiednuscaling}
\end{figure}

%\subsection{Unified constitutive law}
%\label{subsec:unifiedlaw}
%Before seeking the universal constitutive law in vibroconvective turbulence, 
Now, there are two important questions remaining to be answered in vibroconvective turbulence: one is why there exists two different heat transport scaling laws, i.e., $\Nun \sim \Ra_\mathrm{vib}^{1/2}$ and $\Nun \sim \Ra_\mathrm{os}^{1/3}$; the other is whether a unified constitutive law emerges in vibroconvective turbulence. Hereafter, we propose a physical model to address the first question. From the analysis above, we know that the BL-contribution to the global thermal dissipation rate is dominant, implying that the BL dynamics plays a crucial role in heat transport mechanism. In vibroconvective turbulence, there are two types of BL: the thermal boundary layer (TBL) with the thickness of $\delta_\mathrm{th}$, which is estimated by $\delta_\mathrm{th}\approx H/(2\Nun)$; the other is the oscillating boundary layer (OBL) induced by the external vibration. The modulation depth of OBL referring to $\delta_\mathrm{os}$ is defined as the depth, at which the delaying rate of the intensity of vibration-induced shear effect equals to $99\%$. Considering the intensity of vibrational modulation falling off exponentially from the surface, one easily obtains $\delta_\mathrm{os} = -\ln(1-0.99) \delta_S \approx 4.605 \delta_S$ where $\delta_S = \sqrt{2\nu/\Omega}$ is the Stokes layer thickness. First, when $\delta_\mathrm{th} > \delta_\mathrm{os}$ as sketched in figure~\ref{fig:Unifiednuscaling}($a$) above, by taking into account the balance between the convective and conductive transports within TBL, the dimensional analysis of the governing equation of temperature field gives rise to $ w {\Delta}/{\delta_\mathrm{th}} \sim \kappa {\Delta}/{\delta^2_\mathrm{th}}$. And in the momentum equation, the balance between the vibration-induced buoyancy and the viscous dissipation leads to $\alpha A\Omega^2 \Delta \sim \nu {u}/{\delta^2_\mathrm{os}}$ with $u$ the horizontal velocity. Using both above relations, assuming that the magnitude of velocity components $u$ and $w$ follows a similar scaling behavior, i.e., $u\sim w$, together with $\delta_\mathrm{os} \sim \sqrt{\nu/\Omega}$ and $\delta_\mathrm{th} \sim H/\Nun$, one obtains the scaling relation between $\Nun$ and $\Ra_\mathrm{vib}$,
	\begin{equation} \label{eq:nu_ravib}
		\Nun \sim \Ra_\mathrm{vib}^{1/2} \Prn^{1/2}.
	\end{equation}
	The scaling relation in \eqref{eq:nu_ravib} shows that vibroconvective heat transport is independent of viscosity $\nu$, but depends on thermal diffusion coefficient $\kappa$. This implies that the dynamics of TBL is dominant to heat transport in cases of $\delta_\mathrm{th} > \delta_\mathrm{os}$.
	
	When $\delta_\mathrm{th} < \delta_\mathrm{os}$ as sketched in the lower panel of figure~\ref{fig:Unifiednuscaling}($a$), the balance between the vibration-induced buoyancy and the viscous dissipation within TBL allows one to rewrite the momentum equation using dimensional analysis: $\alpha A\Omega^2 \Delta \sim \nu {u}/{\delta^2_\mathrm{th}}$. Combining the above equation and $ w {\Delta}/{\delta_\mathrm{th}} \sim \kappa {\Delta}/{\delta^2_\mathrm{th}}$ for temperature equation, together with  $u\sim w$, $\delta_\mathrm{os} \sim \sqrt{\nu/\Omega}$ and $\delta_\mathrm{th} \sim H/\Nun$, one deduces the scaling relation between $\Nun$ and $\Ra_\mathrm{os}$
	\begin{equation} \label{eq:nu_raos}
		\Nun \sim \Ra_\mathrm{os}^{1/3}.
	\end{equation}
	The heat transport scaling in \eqref{eq:nu_raos} is similar to that of RB convection in the classical regime through replacing the gravitation by vibration-induced acceleration.
	Both heat transport scalings predicted in \eqref{eq:nu_ravib} and \eqref{eq:nu_raos} agree well with numerical results shown in figure~\ref{fig:nuscaling}. The competition between TBL and OBL results in the two different heat transport scaling relations, namely, $\Nun \sim \Ra_\mathrm{vib}^{1/2}$ and $\Nun \sim \Ra_\mathrm{os}^{1/3}$.

	Furthermore, we address the second question of whether the universal constitutive law of vibroconvective turbulence emerges. First, to quantify the dynamics of OBL,  we define the oscillational Reynolds number $\Ren_\mathrm{os} = \alpha \Delta A \Omega \delta_\mathrm{os}/\nu$, which is related to the vibrational velocity with the Boussinesq parameter $\alpha \Delta A \Omega$ and the modulation depth $\delta_\mathrm{os}$. Second, we study the dependency of $\Nun$ on $\Ren_\mathrm{os}$. It is intriguing to find that both $\Nun \sim \Ra_\mathrm{vib}^{1/2}$ and $\Nun \sim \Ra_\mathrm{os}^{1/3}$ scaling laws can be rewritten as $\Nun \sim \Am^{-1} \Ren_\mathrm{os}^\beta$ with $\beta=2$ for the TBL-dominant heat transport regime ($\delta_\mathrm{th}>\delta_\mathrm{os}$), and $\beta=4/3$ for the OBL-dominant heat transport regime ($\delta_\mathrm{th}<\delta_\mathrm{os}$). Therefore, we conclude that due to the competition between the dynamics TBL and OBL on heat transport, the underlying mechanism of heat transport in vibroconvective turbulence can be categorized into two following regimes:
	\begin{enumerate}
		\item [(1)] TBL-dominant regime ($\delta_\mathrm{th} > \delta_\mathrm{os}$): the OBL is submerged into TBL. Thermal plumes facilitated by vibration-induced strong shear detach from OBL and move into TBL. The plume dynamics is then mainly dominant by the molecular diffusion between OBL and TBL. Those plumes thermally diffuse and then self-organize into columnar structures in bulk zones, which transport heat from the bottom hot plate to the top cold one. The heat transport scaling exhibits the scaling $\Nun \sim \Am^{-1} \Ren_\mathrm{os}^{2}$. 
		\item [(2)] OBL-dominant regime ($\delta_\mathrm{os} > \delta_\mathrm{th}$): the TBL is nested into OBL. The OBL dominates the dynamics of thermal plumes ejected from TBL by vibration-induced strong shear. Between OBL and TBL, the shear effect mixes those plumes and sweep away some of them \citep{Scagliarini2014PRE,Blass2020JFM}. The remaining plumes then move into bulk zones and self-organize into columnar structures. In this regime, due to the plume-sweeping mechanism between OBL and TBL, the heat transport is depleted and obeys the scaling with a smaller scaling relation exponent $\Nun \sim \Am^{-1} \Ren_\mathrm{os}^{4/3}$.
	\end{enumerate}
	
	Finally, we use the simulated data to confirm the theoretically deduced unified constitutive law. First, we plot in figures~\ref{fig:Unifiednuscaling}($b,d$) the variation of both $\Am \delta_\mathrm{th}$ and $\Am \delta_\mathrm{os}$ as functions of $\Ren_\mathrm{os}$. It is shown that for all fixed amplitudes, the value of both $\delta_\mathrm{th}$ and $\delta_\mathrm{os}$ monotonically decreases as increasing $\Ren_\mathrm{os}$, and  $\delta_\mathrm{th}$ decreases faster than $\delta_\mathrm{os}$. The insection point between the curves of $\Am \delta_\mathrm{th}$ and $\Am \delta_\mathrm{os}$ divides the the plane into two regions, which corresponds to the TBL-dominant regime in the left side ($\delta_\mathrm{th} > \delta_\mathrm{os}$) and OBL-dominant regime in the right side ($\delta_\mathrm{th} > \delta_\mathrm{os}$). As depicted in the inset of figures~\ref{fig:Unifiednuscaling}($b,d$), the dividing line between TBL-dominant and OBL-dominant regimes is nearly at the position of $\delta_\mathrm{os}/\delta_\mathrm{th} = 1$. This confirms that the underlying mechanism of vibroconvective heat transport is attributed to the competition between the dynamics of TBL and OBL. Second, we plot the calculated $\Am \Nun$ as functions of $\Ren_\mathrm{os}$ as shown in figures~\ref{fig:Unifiednuscaling}($c,e$). It is expected that all numerical data collapse together onto the derived universal constitutive law. Evidently, the numerical data and theoretical model show an excellent agreement. This confirms the emergence of universal constitutive law of vibroconvective turbulence in microgravity.

\section{Conclusions} \label{sec:conclusion}
In summary, we have conducted direct numerical simulations of both 2D and 3D microgravity vibroconvective turbulence over a wide range of dimensionless vibration amplitude and frequency at fixed $\Prn = 4.38$. It is shown that in the absence of gravitational acceleration, vibration creates an ``artificial gravity'' in microgravity to destabilize TBL and trigger massive eruption of thermal plumes. We find that those plumes are finally self-organized into columnar structures in bulk zones to transport heat from the bottom hot plate to the top cold one. This is different from the gravity-induced convection, like Rayleigh-B{\'e}nard convection, in which large-scale circulation is formed in bulk and dominates heat transport. 
By analyzing the basic properties of heat transport, we find the heat transport exhibits two different power-low relations, i.e., $\Nun \sim \Ra_\mathrm{vib}^{1/2}$ at small amplitudes and $\Nun \sim \Ra_\mathrm{os}^{1/3}$ at large amplitudes. Both $\Nun$-relations shows that the global heat flux is independent of the cell height. We also find that the BL-contribution is dominant to the global thermal dissipation rate, implying that the dynamics of boundary layer plays an essential role in vibroconvective heat transport. 
We then propose a physical model to theoretically deduce both $\Nun$-scaling-relations, and explain the distinct properties of viboconvective heat transport, based on the competition between the thermal boundary layer (TBL) and oscillating boundary layer (OBL) induced by the external vibration.  
To look for the universal features, we define the oscillational Reynolds number $\Ren_\mathrm{os}$ quantifying the dynamics of OBL, and study the dependency of heat transport on $\Ren_\mathrm{os}$. Both theoretical results and numerical data shows the emergence of universal constitutive law in vibroconvective turbulence, i.e., $\Nun \sim \Am^{-1} \Ren_\mathrm{os}^{\beta}$, where $\beta$ is the universal scaling exponent. We also find that the exponent $\beta$ is determined by the relative importance between the dynamics of TBL and of OBL to heat transport, and identify $\beta = 2$ in TBL-dominant regime and $\beta = 4/3$ in OBL-dominant regime. It is concluded that the type of vibroconvective turbulence in microgravity owns a universal constitutive law with its underlying heat transport mechanism different from that in gravity-induced convective turbulence. The emergence of universal constitutive laws in vibroconvective turbulence provides a powerful basis on generating a controllable heat transport under microgravity conditions or in microfluidic environment.
% where the effect of gravity is nearly absent, vibroconvective turbulence can provide alternative to effective heat and mass transport in a controllable manner, which surpasses the pure diffusive transport. Looking forward, we expect that our findings can be tested by experimental realizations in space station.

%
\textbf{Acknowledgements:}
This work was supported by the Natural Science Foundation of China under grant nos. 11988102, 11825204, 92052201, 12032016, 12102246, and 11972220. The authors report no conflict of interest.

%\section*{Declaration of interests}
%

%\appendix
%\section{Some derivations}

\bibliographystyle{jfm}
%\bibliography{jfm2esam}
\bibliography{mybibfile}

\begin{thebibliography}{53}
\expandafter\ifx\csname natexlab\endcsname\relax\def\natexlab#1{#1}\fi
\def\au#1{#1} \def\ed#1{#1} \def\yr#1{#1}\def\at#1{#1}\def\jt#1{\textit{#1}}
  \def\bt#1{#1}\def\bvol#1{\textbf{#1}} \def\vol#1{#1} \def\pg#1{#1}
  \def\publ#1{#1}\def\arxiv#1{#1}\def\org#1{#1}\def\st#1{\textit{#1}}

\bibitem[Ahlers {\em et~al.\/}(2022)Ahlers, Bodenschatz, Hartmann, He, Lohse,
  Reiter, Stevens, Verzicco, Wedi, Weiss {\em et~al.\/}]{Ahlers2022aspect}
{\sc \au{Ahlers, G.}, \au{Bodenschatz, E.}, \au{Hartmann, R.}, \au{He, X.},
  \au{Lohse, D.}, \au{Reiter, P.}, \au{Stevens, R.~J.}, \au{Verzicco, R.},
  \au{Wedi, M.}, \au{Weiss, S.} \& \au{others}} \yr{2022}  \at{{Aspect ratio
  dependence of heat transfer in a cylindrical Rayleigh-B{\'e}nard cell}}.
  \jt{Phys.\ Rev.\ Lett.}  \bvol{128}~(8),  \pg{084501}.

\bibitem[Ahlers {\em et~al.\/}(2009)Ahlers, Grossmann \& Lohse]{Ahlers2009RMP}
{\sc \au{Ahlers, G.}, \au{Grossmann, S.} \& \au{Lohse, D.}} \yr{2009}  \at{Heat
  transfer and large scale dynamics in turbulent {Rayleigh--B{\'e}nard}
  convection}.  \jt{Rev.\ Mod.\ Phys.}  \bvol{81}~(2),  \pg{503--537}.

\bibitem[Amiroudine \& Beysens(2008)]{Amiroudine2008PRE}
{\sc \au{Amiroudine, S.} \& \au{Beysens, D.}} \yr{2008}  \at{Thermovibrational
  instability in supercritical fluids under weightlessness}.  \jt{Phys.\ Rev.\
  E}  \bvol{78}~(3),  \pg{036325}.

\bibitem[Apffel {\em et~al.\/}(2021)Apffel, Hidalgo-Caballero, Eddi \&
  Fort]{Apffel2021PNAS}
{\sc \au{Apffel, B.}, \au{Hidalgo-Caballero, S.}, \au{Eddi, A.} \& \au{Fort,
  E.}} \yr{2021}  \at{Liquid walls and interfaces in arbitrary directions
  stabilized by vibrations}.  \jt{Proc.\ Natl.\ Acad.\ Sci.\ U.S.A.}
  \bvol{118}~(48).

\bibitem[Apffel {\em et~al.\/}(2020)Apffel, Novkoski, Eddi \&
  Fort]{Apffel2020nature}
{\sc \au{Apffel, B.}, \au{Novkoski, F.}, \au{Eddi, A.} \& \au{Fort, E.}}
  \yr{2020}  \at{Floating under a levitating liquid}.  \jt{Nature}
  \bvol{585}~(7823),  \pg{48--52}.

\bibitem[Beysens(2006)]{Beysens2006EuroNews}
{\sc \au{Beysens, D.}} \yr{2006}  \at{Vibrations in space as an artificial
  gravity?}  \jt{Europhys. news}  \bvol{37}~(3),  \pg{22--25}.

\bibitem[Beysens {\em et~al.\/}(2005)Beysens, Chatain, Evesque \&
  Garrabos]{Beysens2005PRL}
{\sc \au{Beysens, D.}, \au{Chatain, D.}, \au{Evesque, P.} \& \au{Garrabos, Y.}}
  \yr{2005}  \at{High-frequency driven capillary flows speed up the gas-liquid
  phase transition in zero-gravity conditions}.  \jt{Phys.\ Rev.\ Lett.}
  \bvol{95}~(3),  \pg{034502}.

\bibitem[Blass {\em et~al.\/}(2020)Blass, Zhu, Verzicco, Lohse \&
  Stevens]{Blass2020JFM}
{\sc \au{Blass, A.}, \au{Zhu, X.}, \au{Verzicco, R.}, \au{Lohse, D.} \&
  \au{Stevens, R.~J.}} \yr{2020}  \at{Flow organization and heat transfer in
  turbulent wall sheared thermal convection}.  \jt{J.\ Fluid Mech.}
  \bvol{897},  \pg{A22}.

\bibitem[Brunet {\em et~al.\/}(2007)Brunet, Eggers \& Deegan]{Brunet2007PRL}
{\sc \au{Brunet, P.}, \au{Eggers, J.} \& \au{Deegan, R.~D.}} \yr{2007}
  \at{Vibration-induced climbing of drops}.  \jt{Phys.\ Rev.\ Lett.}
  \bvol{99},  \pg{144501}.

\bibitem[Castaing {\em et~al.\/}(1989)Castaing, Gunaratne, Heslot, Kadanoff,
  Libchaber, Thomae, Wu, Zaleski \& Zanetti]{Castaing1989JFM}
{\sc \au{Castaing, B.}, \au{Gunaratne, G.}, \au{Heslot, F.}, \au{Kadanoff, L.},
  \au{Libchaber, A.}, \au{Thomae, S.}, \au{Wu, X.-Z.}, \au{Zaleski, S.} \&
  \au{Zanetti, G.}} \yr{1989}  \at{{Scaling of hard thermal turbulence in
  Rayleigh-B{\'e}nard convection}}.  \jt{J.\ Fluid Mech.}  \bvol{204},
  \pg{1--30}.

\bibitem[Ciss{\'e} {\em et~al.\/}(2004)Ciss{\'e}, Bardan \&
  Mojtabi]{Cisse2004IJHMT}
{\sc \au{Ciss{\'e}, I.}, \au{Bardan, G.} \& \au{Mojtabi, A.}} \yr{2004}
  \at{{Rayleigh-B{\'e}nard convective instability of a fluid under
  high-frequency vibration}}.  \jt{Int.\ J.\ Heat Mass Tran.}
  \bvol{47}~(19-20),  \pg{4101--4112}.

\bibitem[Crewdson \& Lappa(2021)]{Crewdson2021IJT}
{\sc \au{Crewdson, G.} \& \au{Lappa, M.}} \yr{2021}  \at{Thermally-driven flows
  and turbulence in vibrated liquids}.  \jt{Int. J. Thermofluids}  \bvol{11},
  \pg{100102}.

\bibitem[Daniel {\em et~al.\/}(2005)Daniel, Chaudhury \&
  De~Gennes]{Daniel2005Langmuir}
{\sc \au{Daniel, S.}, \au{Chaudhury, M.~K.} \& \au{De~Gennes, P.-G.}} \yr{2005}
   \at{Vibration-actuated drop motion on surfaces for batch microfluidic
  processes}.  \jt{Langmuir}  \bvol{21}~(9),  \pg{4240--4248}.

\bibitem[Garrabos {\em et~al.\/}(2007)Garrabos, Beysens, Lecoutre, Dejoan,
  Polezhaev \& Emelianov]{Garrabos2007PRE}
{\sc \au{Garrabos, Y.}, \au{Beysens, D.}, \au{Lecoutre, C.}, \au{Dejoan, A.},
  \au{Polezhaev, V.} \& \au{Emelianov, V.}} \yr{2007}  \at{{Thermoconvectional
  phenomena induced by vibrations in supercritical SF$_6$ under
  weightlessness}}.  \jt{Phys.\ Rev.\ E}  \bvol{75}~(5),  \pg{056317}.

\bibitem[Gershuni \& Lyubimov(1998)]{Gershuni1998WS}
{\sc \au{Gershuni, G.~Z.} \& \au{Lyubimov, D.~V.}} \yr{1998} {\em Thermal
  vibrational convection\/}.  \publ{New York: Wiley \& Sons}.

\bibitem[Grossmann \& Lohse(2000)]{Grossmann2000JFM}
{\sc \au{Grossmann, S.} \& \au{Lohse, D.}} \yr{2000}  \at{{Scaling in thermal
  convection: A unifying theory}}.  \jt{J.\ Fluid Mech.}  \bvol{407},
  \pg{27--56}.

\bibitem[Grossmann \& Lohse(2001)]{Grossmann2001PRL}
{\sc \au{Grossmann, S.} \& \au{Lohse, D.}} \yr{2001}  \at{{Thermal convection
  for large Prandtl numbers}}.  \jt{Phys.\ Rev.\ Lett.}  \bvol{86}~(15),
  \pg{3316}.

\bibitem[Guo {\em et~al.\/}(2023)Guo, Wu, Wang, Zhou \& Chong]{guo2023flow}
{\sc \au{Guo, X.}, \au{Wu, J.}, \au{Wang, B.}, \au{Zhou, Q.} \& \au{Chong,
  K.~L.}} \yr{2023}  \at{Flow structure transition in thermal vibrational
  convection}.  \jt{arXiv preprint arXiv:2303.16752} .

\bibitem[He {\em et~al.\/}(2012)He, Funfschilling, Nobach, Bodenschatz \&
  Ahlers]{He2012PRL}
{\sc \au{He, X.}, \au{Funfschilling, D.}, \au{Nobach, H.}, \au{Bodenschatz, E.}
  \& \au{Ahlers, G.}} \yr{2012}  \at{{Transition to the ultimate state of
  turbulent Rayleigh-B{\'e}nard convection}}.  \jt{Phys.\ Rev.\ Lett.}
  \bvol{108}~(2),  \pg{024502}.

\bibitem[Hirata {\em et~al.\/}(2001)Hirata, Sasaki \& Tanigawa]{Hirata2001JFM}
{\sc \au{Hirata, K.}, \au{Sasaki, T.} \& \au{Tanigawa, H.}} \yr{2001}
  \at{Vibrational effects on convection in a square cavity at zero gravity}.
  \jt{J.\ Fluid Mech.}  \bvol{445},  \pg{327--344}.

\bibitem[Iyer {\em et~al.\/}(2020)Iyer, Scheel, Schumacher \&
  Sreenivasan]{Iyer2020PNAS}
{\sc \au{Iyer, K.~P.}, \au{Scheel, J.~D.}, \au{Schumacher, J.} \&
  \au{Sreenivasan, K.~R.}} \yr{2020}  \at{{Classical $1/3$ scaling of
  convection holds up to $Ra=10^{15}$}}.  \jt{Proc.\ Natl.\ Acad.\ Sci.\
  U.S.A.}  \bvol{117}~(14),  \pg{7594--7598}.

\bibitem[Jiang {\em et~al.\/}(2020)Jiang, Zhu, Wang, Huisman \&
  Sun]{Jiang2020SA}
{\sc \au{Jiang, H.}, \au{Zhu, X.}, \au{Wang, D.}, \au{Huisman, S.~G.} \&
  \au{Sun, C.}} \yr{2020}  \at{Supergravitational turbulent thermal
  convection}.  \jt{Sci.\ Adv.}  \bvol{6}~(40),  \pg{eabb8676}.

\bibitem[Kraichnan(1962)]{Kraichnan1962PoF}
{\sc \au{Kraichnan, R.~H.}} \yr{1962}  \at{{Turbulent thermal convection at
  arbitrary Prandtl number}}.  \jt{Phys. Fluids}  \bvol{5}~(11),
  \pg{1374--1389}.

\bibitem[Lepot {\em et~al.\/}(2018)Lepot, Auma{\^\i}tre \&
  Gallet]{Lepot2018PNAS}
{\sc \au{Lepot, S.}, \au{Auma{\^\i}tre, S.} \& \au{Gallet, B.}} \yr{2018}
  \at{Radiative heating achieves the ultimate regime of thermal convection}.
  \jt{Proc.\ Natl.\ Acad.\ Sci.\ U.S.A.}  \bvol{115}~(36),  \pg{8937--8941}.

\bibitem[Lohse \& Xia(2010)]{Lohse2010ARFM}
{\sc \au{Lohse, D.} \& \au{Xia, K.~Q.}} \yr{2010}  \at{Small-scale properties
  of turbulent {Rayleigh--B{\'e}nard} convection}.  \jt{Annu.\ Rev.\ Fluid
  Mech.}  \bvol{42}~(1),  \pg{335--364}.

\bibitem[Malkus(1954)]{Malkus1954PRSLA}
{\sc \au{Malkus, W.~V.}} \yr{1954}  \at{The heat transport and spectrum of
  thermal turbulence}.  \jt{Proc. R. Soc. Lond. A}  \bvol{225}~(1161),
  \pg{196--212}.

\bibitem[Mialdun {\em et~al.\/}(2008)Mialdun, Ryzhkov, Melnikov \&
  Shevtsova]{Mialdun2008PRL}
{\sc \au{Mialdun, A.}, \au{Ryzhkov, I.}, \au{Melnikov, D.} \& \au{Shevtsova,
  V.}} \yr{2008}  \at{Experimental evidence of thermal vibrational convection
  in a nonuniformly heated fluid in a reduced gravity environment}.  \jt{Phys.\
  Rev.\ Lett.}  \bvol{101}~(8),  \pg{084501}.

\bibitem[Monti {\em et~al.\/}(2001)Monti, Savino \& Lappa]{Monti2001AA}
{\sc \au{Monti, R.}, \au{Savino, R.} \& \au{Lappa, M.}} \yr{2001}  \at{On the
  convective disturbances induced by g-jitter on the space station}.  \jt{Acta
  Astronaut.}  \bvol{48}~(5-12),  \pg{603--615}.

\bibitem[Niemela {\em et~al.\/}(2000)Niemela, Skrbek, Sreenivasan \&
  Donnelly]{Niemela2000Nature}
{\sc \au{Niemela, J.}, \au{Skrbek, L.}, \au{Sreenivasan, K.} \& \au{Donnelly,
  R.}} \yr{2000}  \at{{Turbulent convection at very high Rayleigh numbers}}.
  \jt{Nature}  \bvol{404}~(6780),  \pg{837--840}.

\bibitem[Pesch {\em et~al.\/}(2008)Pesch, Palaniappan, Tao \&
  Busse]{Pesch2008JFM}
{\sc \au{Pesch, W.}, \au{Palaniappan, D.}, \au{Tao, J.} \& \au{Busse, F.~H.}}
  \yr{2008}  \at{Convection in heated fluid layers subjected to time-periodic
  horizontal accelerations}.  \jt{J.\ Fluid Mech.}  \bvol{596},  \pg{313--332}.

\bibitem[Plumley \& Julien(2019)]{Plumley2019Earth}
{\sc \au{Plumley, M.} \& \au{Julien, K.}} \yr{2019}  \at{{Scaling laws in
  Rayleigh-B{\'e}nard convection}}.  \jt{Earth Space Sci.}  \bvol{6}~(9),
  \pg{1580--1592}.

\bibitem[Priestley(1954)]{Priestley1954AJP}
{\sc \au{Priestley, C.}} \yr{1954}  \at{Convection from a large horizontal
  surface}.  \jt{Aust. J. Phys.}  \bvol{7}~(1),  \pg{176--201}.

\bibitem[Rogers {\em et~al.\/}(2000{\natexlab{{\em a\/}}})Rogers, Schatz,
  Bougie \& Swift]{Rogers2000PRL}
{\sc \au{Rogers, J.~L.}, \au{Schatz, M.~F.}, \au{Bougie, J.~L.} \& \au{Swift,
  J.~B.}} \yr{2000{\natexlab{{\em a\/}}}}  \at{{Rayleigh-B{\'e}nard convection
  in a vertically oscillated fluid layer}}.  \jt{Phys.\ Rev.\ Lett.}
  \bvol{84}~(1),  \pg{87}.

\bibitem[Rogers {\em et~al.\/}(2000{\natexlab{{\em b\/}}})Rogers, Schatz,
  Brausch \& Pesch]{Rogers2000PRLb}
{\sc \au{Rogers, J.~L.}, \au{Schatz, M.~F.}, \au{Brausch, O.} \& \au{Pesch,
  W.}} \yr{2000{\natexlab{{\em b\/}}}}  \at{{Superlattice patterns in
  vertically oscillated Rayleigh-B{\'e}nard convection}}.  \jt{Phys.\ Rev.\
  Lett.}  \bvol{85}~(20),  \pg{4281}.

\bibitem[Salgado~S\'anchez {\em et~al.\/}(2019)Salgado~S\'anchez, Gaponenko,
  Porter \& Shevtsova]{Salgado2019PRE}
{\sc \au{Salgado~S\'anchez, P.}, \au{Gaponenko, Y.~A.}, \au{Porter, J.} \&
  \au{Shevtsova, V.}} \yr{2019}  \at{Finite-size effects on pattern selection
  in immiscible fluids subjected to horizontal vibrations in weightlessness}.
  \jt{Phys. Rev. E}  \bvol{99},  \pg{042803}.

\bibitem[S{\'a}nchez {\em et~al.\/}(2020)S{\'a}nchez, Gaponenko, Yasnou,
  Mialdun, Porter \& Shevtsova]{Sanchez2020JFM}
{\sc \au{S{\'a}nchez, P.~S.}, \au{Gaponenko, Y.}, \au{Yasnou, V.}, \au{Mialdun,
  A.}, \au{Porter, J.} \& \au{Shevtsova, V.}} \yr{2020}  \at{Effect of initial
  interface orientation on patterns produced by vibrational forcing in
  microgravity}.  \jt{J.\ Fluid Mech.}  \bvol{884}.

\bibitem[S{\'a}nchez {\em et~al.\/}(2019)S{\'a}nchez, Yasnou, Gaponenko,
  Mialdun, Porter \& Shevtsova]{Sanchez2019JFM}
{\sc \au{S{\'a}nchez, P.~S.}, \au{Yasnou, V.}, \au{Gaponenko, Y.}, \au{Mialdun,
  A.}, \au{Porter, J.} \& \au{Shevtsova, V.}} \yr{2019}  \at{Interfacial
  phenomena in immiscible liquids subjected to vibrations in microgravity}.
  \jt{J.\ Fluid Mech.}  \bvol{865},  \pg{850--883}.

\bibitem[Savino {\em et~al.\/}(1998)Savino, Monti \& Piccirillo]{Savino1998CF}
{\sc \au{Savino, R.}, \au{Monti, R.} \& \au{Piccirillo, M.}} \yr{1998}
  \at{Thermovibrational convection in a fluid cell}.  \jt{Comput. Fluids}
  \bvol{27}~(8),  \pg{923--939}.

\bibitem[Scagliarini {\em et~al.\/}(2014)Scagliarini, Gylfason \&
  Toschi]{Scagliarini2014PRE}
{\sc \au{Scagliarini, A.}, \au{Gylfason, {\'A}.} \& \au{Toschi, F.}} \yr{2014}
  \at{{Heat-flux scaling in turbulent Rayleigh-B{\'e}nard convection with an
  imposed longitudinal wind}}.  \jt{Phys.\ Rev.\ E}  \bvol{89}~(4),
  \pg{043012}.

\bibitem[Sharma {\em et~al.\/}(2019)Sharma, Erriguible, Gandikota, Beysens \&
  Amiroudine]{Sharma2019PRF}
{\sc \au{Sharma, D.}, \au{Erriguible, A.}, \au{Gandikota, G.}, \au{Beysens, D.}
  \& \au{Amiroudine, S.}} \yr{2019}  \at{Vibration-induced thermal
  instabilities in supercritical fluids in the absence of gravity}.  \jt{Phys.\
  Rev.\ Fluids}  \bvol{4}~(3),  \pg{033401}.

\bibitem[Shevtsova {\em et~al.\/}(2010)Shevtsova, Ryzhkov, Melnikov, Gaponenko
  \& Mialdun]{Shevtsova2010JFM}
{\sc \au{Shevtsova, V.}, \au{Ryzhkov, I.~I.}, \au{Melnikov, D.~E.},
  \au{Gaponenko, Y.~A.} \& \au{Mialdun, A.}} \yr{2010}  \at{Experimental and
  theoretical study of vibration-induced thermal convection in low gravity}.
  \jt{J.\ Fluid Mech.}  \bvol{648},  \pg{53--82}.

\bibitem[Spiegel(1963)]{Spiegel1963AstJ}
{\sc \au{Spiegel, E.~A.}} \yr{1963}  \at{A generalization of the mixing-length
  theory of turbulent convection}.  \jt{Astrophys. J.}  \bvol{138},  \pg{216}.

\bibitem[Sreenivasan(2019)]{Sreenivasan2019PNAS}
{\sc \au{Sreenivasan, K.~R.}} \yr{2019}  \at{{Turbulent mixing: A
  perspective}}.  \jt{Proc.\ Natl.\ Acad.\ Sci.\ U.S.A.}  \bvol{116}~(37),
  \pg{18175--18183}.

\bibitem[Stevens {\em et~al.\/}(2013)Stevens, van~der Poel, Grossmann \&
  Lohse]{Stevens2013JFM}
{\sc \au{Stevens, R.~J.}, \au{van~der Poel, E.~P.}, \au{Grossmann, S.} \&
  \au{Lohse, D.}} \yr{2013}  \at{The unifying theory of scaling in thermal
  convection: the updated prefactors}.  \jt{J.\ Fluid Mech.}  \bvol{730},
  \pg{295--308}.

\bibitem[Swaminathan {\em et~al.\/}(2018)Swaminathan, Garrett, Poese \&
  Smith]{Swaminathan2018JASA}
{\sc \au{Swaminathan, A.}, \au{Garrett, S.~L.}, \au{Poese, M.~E.} \& \au{Smith,
  R.~W.}} \yr{2018}  \at{{Dynamic stabilization of the Rayleigh-B{\'e}nard
  instability by acceleration modulation}}.  \jt{J. Acoust. Soc. Am.}
  \bvol{144}~(4),  \pg{2334--2343}.

\bibitem[Toppaladoddi {\em et~al.\/}(2017)Toppaladoddi, Succi \&
  Wettlaufer]{Toppaladoddi2017PRL}
{\sc \au{Toppaladoddi, S.}, \au{Succi, S.} \& \au{Wettlaufer, J.~S.}} \yr{2017}
   \at{Roughness as a route to the ultimate regime of thermal convection}.
  \jt{Phys.\ Rev.\ Lett.}  \bvol{118}~(7),  \pg{074503}.

\bibitem[Wang {\em et~al.\/}(2020)Wang, Zhou \& Sun]{Wang2020SA}
{\sc \au{Wang, B.-F.}, \au{Zhou, Q.} \& \au{Sun, C.}} \yr{2020}
  \at{Vibration-induced boundary-layer destabilization achieves massive
  heat-transport enhancement}.  \jt{Sci.\ Adv.}  \bvol{6}~(21),  \pg{eaaz8239}.

\bibitem[Wang {\em et~al.\/}(2019)Wang, Mathai \& Sun]{Wang2019NC}
{\sc \au{Wang, Z.}, \au{Mathai, V.} \& \au{Sun, C.}} \yr{2019}
  \at{Self-sustained biphasic catalytic particle turbulence}.  \jt{Nat.\
  Commun.}  \bvol{10},  \pg{3333}.

\bibitem[Wu {\em et~al.\/}(2021)Wu, Dong, Wang \& Zhou]{Wu2021PoF}
{\sc \au{Wu, J.-Z.}, \au{Dong, Y.-H.}, \au{Wang, B.-F.} \& \au{Zhou, Q.}}
  \yr{2021}  \at{{Phase decomposition analysis on oscillatory
  Rayleigh--B{\'e}nard turbulence}}.  \jt{Phys.\ Fluids}  \bvol{33}~(4),
  \pg{045108}.

\bibitem[Wu {\em et~al.\/}(2022)Wu, Wang, Chong, Dong, Sun \& Zhou]{wu2022jfm}
{\sc \au{Wu, J.-Z.}, \au{Wang, B.-F.}, \au{Chong, K.~L.}, \au{Dong, Y.-H.},
  \au{Sun, C.} \& \au{Zhou, Q.}} \yr{2022}  \at{{Vibration-induced
  `anti-gravity' tames thermal turbulence at high Rayleigh numbers}}.  \jt{J.\
  Fluid Mech.}  \bvol{951},  \pg{A13}.

\bibitem[Zhu {\em et~al.\/}(2018)Zhu, Verschoof, Bakhuis, Huisman, Verzicco,
  Sun \& Lohse]{Zhu2018NP}
{\sc \au{Zhu, X.}, \au{Verschoof, R.~A.}, \au{Bakhuis, D.}, \au{Huisman,
  S.~G.}, \au{Verzicco, R.}, \au{Sun, C.} \& \au{Lohse, D.}} \yr{2018}
  \at{Wall roughness induces asymptotic ultimate turbulence}.  \jt{Nat.\ Phys.}
   \bvol{14}~(4),  \pg{417--423}.

\bibitem[Zou \& Yang(2021)]{zou2021jfm}
{\sc \au{Zou, S.} \& \au{Yang, Y.}} \yr{2021}  \at{Realizing the ultimate
  scaling in convection turbulence by spatially decoupling the thermal and
  viscous boundary layers}.  \jt{J.\ Fluid Mech.}  \bvol{919},  \pg{R3}.

\bibitem[Zyuzgin {\em et~al.\/}(2001)Zyuzgin, Ivanov, Polezhaev, Putin \&
  Soboleva]{Zyuzgin2001CS}
{\sc \au{Zyuzgin, A.}, \au{Ivanov, A.}, \au{Polezhaev, V.}, \au{Putin, G.} \&
  \au{Soboleva, E.}} \yr{2001}  \at{Convective motions in near-critical fluids
  under real zero-gravity conditions}.  \jt{Cosmic Res.}  \bvol{39}~(2),
  \pg{175--186}.

\end{thebibliography}

%% End of file `jfm2esam.bib'.

\end{document}